\PassOptionsToPackage{unicode}{hyperref}
\PassOptionsToPackage{hyphens}{url}
\PassOptionsToPackage{dvipsnames,svgnames,x11names}{xcolor}
\documentclass[
]{article}
\usepackage{amsmath,amssymb}
\usepackage{lmodern}
\usepackage{iftex}
\ifPDFTeX
  \usepackage[T1]{fontenc}
  \usepackage[utf8]{inputenc}
  \usepackage{textcomp} 
\else 
  \usepackage{unicode-math}
  \defaultfontfeatures{Scale=MatchLowercase}
  \defaultfontfeatures[\rmfamily]{Ligatures=TeX,Scale=1}
\fi
\IfFileExists{upquote.sty}{\usepackage{upquote}}{}
\IfFileExists{microtype.sty}{
  \usepackage[]{microtype}
  \UseMicrotypeSet[protrusion]{basicmath} 
}{}
\makeatletter
\@ifundefined{KOMAClassName}{
  \IfFileExists{parskip.sty}{%
    \usepackage{parskip}
  }{
    \setlength{\parindent}{0pt}
    \setlength{\parskip}{6pt plus 2pt minus 1pt}}
}{
  \KOMAoptions{parskip=half}}
\makeatother
\usepackage{xcolor}
\usepackage{longtable,booktabs,array}
\usepackage{calc} 
\usepackage{etoolbox}
\makeatletter
\patchcmd\longtable{\par}{\if@noskipsec\mbox{}\fi\par}{}{}
\makeatother
\IfFileExists{footnotehyper.sty}{\usepackage{footnotehyper}}{\usepackage{footnote}}
\makesavenoteenv{longtable}
\usepackage{graphicx}
\makeatletter
\def\maxwidth{\ifdim\Gin@nat@width>\linewidth\linewidth\else\Gin@nat@width\fi}
\def\maxheight{\ifdim\Gin@nat@height>\textheight\textheight\else\Gin@nat@height\fi}
\makeatother
\setkeys{Gin}{width=\maxwidth,height=\maxheight,keepaspectratio}
\makeatletter
\def\fps@figure{htbp}
\makeatother
\providecommand{\tightlist}{%
  \setlength{\itemsep}{0pt}\setlength{\parskip}{0pt}}
\newlength{\cslhangindent}
\newlength{\csllabelwidth}
\newlength{\cslentryspacingunit} 
\newenvironment{CSLReferences}[2] 
 {
  \setlength{\parindent}{0pt}
  \ifodd #1
  \let\oldpar\par
  \def\par{\hangindent=\cslhangindent\oldpar}
  \fi
  \setlength{\parskip}{#2\cslentryspacingunit}
 }%
 {}
\usepackage{calc}

\ifLuaTeX
\usepackage[bidi=basic]{babel}
\else
\usepackage[bidi=default]{babel}
\fi
\babelprovide[main,import]{american}

\def\languageshorthands#1{}
\ifLuaTeX
  \usepackage{selnolig}  
\fi
\IfFileExists{bookmark.sty}{\usepackage{bookmark}}{\usepackage{hyperref}}
\IfFileExists{xurl.sty}{\usepackage{xurl}}{} 
\hypersetup{
  pdftitle={Zero-Shot Duet Singing Voices Separation with Diffusion
Models},
  pdfauthor={Chin-Yun Yu, Emilian Postolache, Emanuele Rodolà, György
Fazekas},
  pdflang={en-US},
  colorlinks=true,
  linkcolor={Maroon},
  filecolor={Maroon},
  citecolor={Blue},
  urlcolor={Blue},
  pdfcreator={LaTeX via pandoc}}

\title{Zero-Shot Duet Singing Voices Separation with Diffusion Models}

\usepackage[affil-it]{authblk}
\usepackage{orcidlink}
\author[1%
  ]{Chin-Yun Yu%
    \,\orcidlink{0000-0003-3782-2713}\,%
    }
\author[2%
  ]{Emilian Postolache%
    }
\author[2%
  ]{Emanuele Rodolà%
    }
\author[1%
  ]{György Fazekas%
    }

\affil[1]{Queen Mary University of London}
\affil[2]{Sapienza University of Rome}
\date{4 November 2023}

\begin{document}
\maketitle

\hypertarget{abstract}{%
\section{Abstract}\label{abstract}}

In recent studies, diffusion models have shown promise as priors for
solving audio inverse problems
(\protect\hyperlink{ref-moliner2023solving}{Moliner et al., 2023};
\protect\hyperlink{ref-murata2023gibbsddrm}{Murata et al., 2023};
\protect\hyperlink{ref-saito2023unsupervised}{Saito et al., 2023};
\protect\hyperlink{ref-yu2023conditioning}{Yu et al., 2023}), including
source separation (\protect\hyperlink{ref-mariani2023multi}{Mariani et
al., 2023}). These models allow us to sample from the posterior
distribution of a target signal given an observed signal by manipulating
the diffusion process. However, when separating audio sources of the
same type, such as duet singing voices, the prior learned by the
diffusion process may not be sufficient to maintain the consistency of
the source identity in the separated audio. For example, the singer may
change from one to another from time to time. Tackling this problem will
be useful for separating sources in a choir, or a mixture of multiple
instruments with similar timbre, without acquiring large amounts of
paired data. In this paper, we examine this problem in the context of
duet singing voices separation, and propose a method to enforce the
coherency of singer identity by splitting the mixture into overlapping
segments and performing posterior sampling in an auto-regressive manner,
conditioning on the previous segment. We evaluate the proposed method on
the MedleyVox dataset (\protect\hyperlink{ref-medleyvox}{Jeon et al.,
2023}) with different overlap ratios, and show that the proposed method
outperforms naive posterior sampling baseline. Our source code and the
pre-trained model are publicly available on
\url{https://github.com/iamycy/duet-svs-diffusion}.

\hypertarget{introduction}{%
\section{Introduction}\label{introduction}}

With the success of audio generative models, such as OpenAI's Jukebox
(\protect\hyperlink{ref-jukebox}{Dhariwal et al., 2020}), works have
been conducted to leverage them in unsupervised source separation
(\protect\hyperlink{ref-manilow2021unsupervised}{Manilow et al., 2021};
\protect\hyperlink{ref-zai2022transfer}{Zai El Amri et al., 2022}). The
benefit of this approach is that the training is done on isolated
sources and does not require paired multi-track data, which is sometimes
difficult to obtain. Moreover, the rise of diffusion models (DMs)
(\protect\hyperlink{ref-ddpm}{Ho et al., 2020};
\protect\hyperlink{ref-song2020score}{Song et al., 2020}) introduce a
more flexible way for unsupervised source separation compared to the
traditional generative models.

By adding conditional information into the diffusion process, we can
sample the target signal from a pre-trained unconditional DM given an
observed signal. This process is called posterior sampling, and has been
applied successfully in solving various audio inverse problems
(\protect\hyperlink{ref-vrdmg}{Hernandez-Olivan et al., 2023};
\protect\hyperlink{ref-moliner2023solving}{Moliner et al., 2023};
\protect\hyperlink{ref-murata2023gibbsddrm}{Murata et al., 2023};
\protect\hyperlink{ref-saito2023unsupervised}{Saito et al., 2023};
\protect\hyperlink{ref-yu2023conditioning}{Yu et al., 2023}) and also
source separation(\protect\hyperlink{ref-hirano2023diffusion}{Hirano et
al., 2023}; \protect\hyperlink{ref-mariani2023multi}{Mariani et al.,
2023}). Mariani et al. (\protect\hyperlink{ref-mariani2023multi}{2023})
trained a 4-track DM, each track correponds to \(Bass\), \(Drums\),
\(Guitar\), and \(Piano\), and proposed a novel conditioning scheme
based on the Dirac delta function to do music source separation with
posterior sampling. Hirano et al.
(\protect\hyperlink{ref-hirano2023diffusion}{2023}) proposed to use an
unconditional speech DM to enhance an initial estimation of a
multi-speaker speech separation model. They assume the target speech is
drawn from a Gaussian distribution centered at the initial estimation
and use Diffusion Denoising Restoration Models
(\protect\hyperlink{ref-ddrm}{Kawar et al., 2022}) for refining the
estimation. Nevertheless, no similar work has proposed for source
separation of monotimbral sources, such as singing voices, solely based
on a single unconditional DM.

We start examining this problem on singing voices separation with an
unconditional single singer DM and found that, without additional
guidance similar to Hirano et al.
(\protect\hyperlink{ref-hirano2023diffusion}{2023}), the singer identity
in the separated audio is not consistent and can switch from one to
another after a short period of time. This is reasonable because the
prior we can use are the implicit timbre conherency and the pitch
contour distribution learned by the DMs. Moreover, in case the sources
in the mixture are sung by the same singer, which is common in studio
recordings, only the pitch prior can be utilised. Interestingly, this
kind of failure cases also exist in supervised separator
(\protect\hyperlink{ref-medleyvox}{Jeon et al., 2023}), showing that the
problem is not trivial.

In this paper, we propose to tackle this problem by splitting the
mixture into overlapping segments and perform posterior sampling
sequentially. The mixture of the segment and the overlapping part of the
previous segment are used as condition. We hypothesise that conditioning
on the previous segment can guide the posterior sampling to maintain the
singer identity. In addition, decomposing the distribution into a chain
of conditional distributions also gives users finer control over the
separation process, suitable for human-in-the-loop applications.

\hypertarget{problem-formulation}{%
\section{Problem Formulation}\label{problem-formulation}}

Let us start with a general formulation of the multi-channel audio
inverse problem. We have an M-channel mixture signal
\(\mathbf{x}(n, f) \in \mathbb{C}^M\) where \(n, f\) are the time and
frequency indices, respectively. It contains \(N\) sources
\(s_i(n, f) \in \mathbb{C}^N\) where \(i \in \{1, 2, ..., N\}\), and a
measurement noise signal \(z \sim \mathcal{CN}(0, \sigma_x^2)\). The
sources are transformed by a non-invertible linear system
\(\mathbf{H}(n, f) \in \mathbb{C}^{M \times N}\) and mixed with the
noise, resulting in \[
\mathbf{x}(n, f) = \mathbf{H}(n, f)
\begin{bmatrix}
s_1(n, f) \\
s_2(n, f) \\
\vdots \\
s_N(n, f)
\end{bmatrix}
+ z.
\label{mix}
\] The objective of the inverse problem is to estimate the sources
\(\mathbf{s}(n, f) \in \mathbb{C}^N\) from the mixture
\(\mathbf{x}(n, f)\). To solve this with posterior sampling, we need to
train DMs that model the distribution of the sources \(p(s_i(n, f))\).
Each \(s_i(n, f)\) is either drawn from different or the same
distributions (DMs), where the latter is the case we want to tackle in
this paper. Note that by distribution we mean the sounds sharing very
similar timbre, such as singing voices or string instruments.

\hypertarget{related-works}{%
\subsection{Related works}\label{related-works}}

The work by Mariani et al.
(\protect\hyperlink{ref-mariani2023multi}{2023}) is one of the pioneer
that use posterior sampling to do source separation. They consider
single channel source separation where \(\mathbf{H}(n, f)\) is simply an
all-one vector. However, the authors modelled the joint distribution
\(p(s_1(n, f), s_2(n, f), \cdots, s_N(n, f))\) with a single DM and
utilise the inter-source correlation to do separation. The
joint-training method also cannot generalise to an arbitrary number of
sources. The work by Hirano et al.
(\protect\hyperlink{ref-hirano2023diffusion}{2023}) is the closest work
to ours. The DM they used is trained on single speaker speech data, but
the requirement of a pre-trained speech separation model breaks the
fully unsupervised assumption.

Outside source separation, several works have dealt with this problem,
either with a known \(\mathbf{H}(n, f)\) in the case of bandwidth
extension (\protect\hyperlink{ref-vrdmg}{Hernandez-Olivan et al., 2023};
\protect\hyperlink{ref-moliner2023solving}{Moliner et al., 2023};
\protect\hyperlink{ref-yu2023conditioning}{Yu et al., 2023}) or an
unknown one such as removing reverberation from vocals
(\protect\hyperlink{ref-murata2023gibbsddrm}{Murata et al., 2023};
\protect\hyperlink{ref-saito2023unsupervised}{Saito et al., 2023}).
Non-linear problems such as de-cliping has also been tackled with DMs
(\protect\hyperlink{ref-vrdmg}{Hernandez-Olivan et al., 2023};
\protect\hyperlink{ref-moliner2023solving}{Moliner et al., 2023}).

Lastly, we point out that, there are no works, to the best of our
knowledge, that use this approach to solve problems with multiple
sources and multi-channel mixtures. This problem occurs in ensemble
separation, such as choir or orchestra sections, which is often recorded
with multiple microphones. The holy grail of this approach is to have a
generalised solution that can be applied on arbitrary transformation
\(\mathbf{H}(n, f)\) and \(N\) with the same set of DMs, and each of the
DMs is trained on isolated sources without the need of mixture data.

\hypertarget{methodology}{%
\section{Methodology}\label{methodology}}

In diffusion models, the data generation process is governed by an ODE
\[
d\mathbf{s}(t) = \sigma(t) \nabla_{\mathbf{s}(t)} \log p(\mathbf{s}(t)) dt.
\label{ode}
\] Here, we use \(\mathbf{s}(t)\) to represent
\(s_i(n, f) + z, z \sim N(0, \sigma^2(t))\) for arbitrary \(i\). The
\(\sigma(t)\) is a increasing function of \(t\) and is called the noise
schedule. Because \(\mathbf{s}(0) \approx s_i(n, f)\), we can sample
them by integrating the ODE from \(t=T\) to \(t=0\) with
\(\mathbf{s}(T) \sim N(\mathbf{0}, \sigma^2(T))\). We can use a neural
network \(\theta(\mathbf{s}(t); t)\) to estimate the score function
\(\nabla_{\mathbf{s}(t)} \log p(\mathbf{s}(t))\) by training them to
denoise the noisy data \(\mathbf{s}(t)\), where \(t \sim U(0, T)\).

\hypertarget{mixture-conditioning}{%
\subsection{Mixture conditioning}\label{mixture-conditioning}}

One can transform the score function to a conditional one using simple
bayes rule \[
\nabla_{\mathbf{s}(t)} \log p(\mathbf{s}(t)|\mathbf{x}) = \nabla_{\mathbf{s}(t)} \log p(\mathbf{x}|\mathbf{s}(t)) + \nabla_{\mathbf{s}(t)} \log p(\mathbf{s}(t)).
\label{cond}
\] We consider the case where the mixture is simply a sum of the
sources, i.e.~\(\mathbf{x} = \sum_{i = 1}^N \mathbf{s}_i(0)\). We choose
the weakly-supervised posterior score function from Mariani et al.
(\protect\hyperlink{ref-mariani2023multi}{2023}) as our conditional
score function, which is

\[
\nabla_{\mathbf{s}_i(t)} \log p(\mathbf{s}_i(t)|\mathbf{x}) \approx
\nabla_{\mathbf{s}_i(t)} \log p(\mathbf{s}_i(t)) - 
\nabla_{\mathbf{s}_i(t)} \log p(\mathbf{x} - \sum_{i = 2}^N \mathbf{s}_i(t))
\label{cond2}
\] for \(i > 1\) and we set \(\mathbf{s}_1(t)\) as the constrained
source.

\hypertarget{enforcing-coherency-with-auto-regressive-inpainting}{%
\subsection{Enforcing coherency with auto-regressive
inpainting}\label{enforcing-coherency-with-auto-regressive-inpainting}}

\begin{figure}
\centering
\includegraphics[width=0.8\textwidth,height=\textheight]{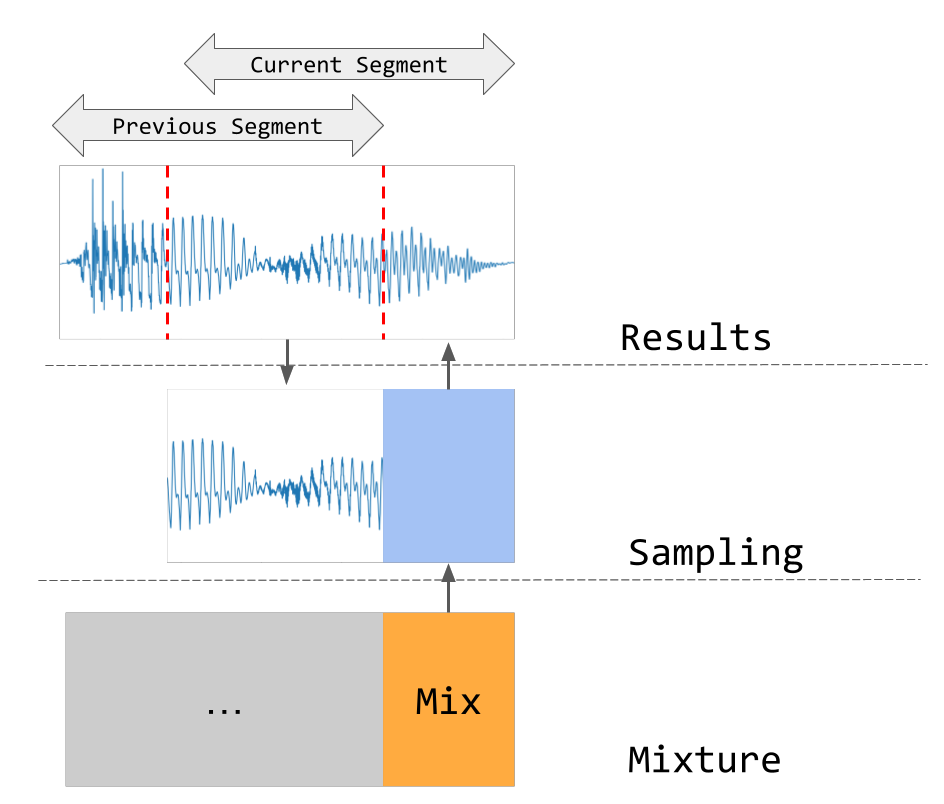}
\caption{A diagram of the conditioning process on two consecutive
segments.\label{fig:diagram}}
\end{figure}

To tackle the problem of singer identity switching, we propose to split
the mixture into overlapping segments and perform posterior sampling
sequentially. The mixture of the segment and the overlapping part of the
previous separated segment are used as condition. The conditioning is
simply placing the noisy \(\mathbf{s}(t)\) condition on the overlapping
part during sampling, similar to inpainting
(\protect\hyperlink{ref-crash}{Rouard \& Hadjeres, 2021}). This
regulates the model to predict more coherent signal to the condition in
the non-overlapping part. The whole sampling process is illustrated in
Figure \ref{fig:diagram}.

\hypertarget{experiments}{%
\section{Experiments}\label{experiments}}

\hypertarget{datasets}{%
\subsection{Datasets}\label{datasets}}

We used the following 8 singing voice datasets for training our
unconditional DM. The total duration of the training data is
\textgreater{} 104 hours.

\begin{itemize}
\tightlist
\item
  M4Singer (\protect\hyperlink{ref-m4singer}{Zhang et al., 2022})
\item
  OpenCPOP (\protect\hyperlink{ref-opencpop}{Wang et al., 2022})
\item
  OpenSinger (\protect\hyperlink{ref-opensinger}{Huang et al., 2021})
\item
  VocalSet (\protect\hyperlink{ref-vocalset}{Wilkins et al., 2018})
\item
  JVS-MuSiC (\protect\hyperlink{ref-jvs}{Tamaru et al., 2020})
\item
  Children's Song Dataset (\protect\hyperlink{ref-csd}{Choi et al.,
  2020})
\item
  NUS Sung and Spoken Lyrics Corpus (\protect\hyperlink{ref-nus}{Duan et
  al., 2013})
\item
  PJS (\protect\hyperlink{ref-pjs}{Koguchi et al., 2020})
\end{itemize}

We resampled all the data to 24 kHz and converted them to mono and
segmented the data into 131072 samples (5.46 seconds) with half of the
samples overlapping. The test songs are from the duet category of
MedleyVox dataset (\protect\hyperlink{ref-medleyvox}{Jeon et al.,
2023}). We dropped 13 clips with loud background music or have effects
such as reverb and echo\footnote{We dropped clips from the songs
  \emph{CatMartino\_IPromise}, \emph{TleilaxEnsemble\_Late}, and
  \emph{TleilaxEnsemble\_MelancholyFlowers}.}, resulting in 103 clips
(also resampled to 24 kHz) for evaluation.

\hypertarget{trainingevaluation-details}{%
\subsection{Training/Evaluation
details}\label{trainingevaluation-details}}

The score prediction network is a UNet developed by Schneider et al.
(\protect\hyperlink{ref-mousai}{2023}) with 424M parameters. We
recommend readers to go to our code repository for more details on the
configurations. We used AdamW (\protect\hyperlink{ref-adamw}{Loshchilov
\& Hutter, 2017}) with a learning rate of 0.0001 and a batch size of 32.
We trained it for 1M steps, which is roungly 8 days on a single RTX
A5000 GPU. The evaluation metrics we used are SI-SDR and SDR
improvements (SI-SDRi and SDRi) over the mixture. We calculated the
metrics with the \texttt{asteroid}
(\protect\hyperlink{ref-asteroid}{Pariente et al., 2020}) package. The
number of diffusion steps were set to 100 for all the experiments. The
segment size for auto-regressive sampling was the same as the training
data, i.e.~131072 samples. We tested different overlap ratios for
sampling. Generall speaking, the more overlap the better the results,
but the sampling time also increases. We choose 75\% overlap as a good
trade-off between the performance and the sampling time.

\hypertarget{baselines}{%
\subsection{Baselines}\label{baselines}}

We evaluated the following methods:

\begin{itemize}
\tightlist
\item
  Non-negative Matrix Factorisation (\textbf{NMF}): We reproduced the
  NMF baseline from Schulze-Forster et al.
  (\protect\hyperlink{ref-schulze2023unsupervised}{2023}) using
  \texttt{torchnmf}\footnote{\url{https://github.com/yoyololicon/pytorch-NMF}}. A modified version of the multi-pitch estimator from Yu \& Su
  (\protect\hyperlink{ref-yu2018multi}{2018}) was used to estimate
  candidate pitches for initialising the activation matrix.
\item
  \textbf{Naive}: The whole mixture is used as condition without
  segmentation and auto-regressive sampling. We sampled each mixture 3
  times and pick the lowest loss one to the ground truth.
\item
  \textbf{AR}: The proposed auto-regressive sampling method. We sampled
  three times for each segment and pick the lowest loss one to the
  ground truth before move on to the next segment.
\item
  \textbf{Segmented}: The mixture is segmented into 131072 * 25\% =
  32768 samples without overlapping and naive sampling is performed on
  each segment.
\end{itemize}

\hypertarget{results-and-discussions}{%
\subsection{Results and discussions}\label{results-and-discussions}}

For all the posterior sampling methods, we ran them three times and
report the mean and standard deviation of the metrics in the following
table. The first row is the supervised separator scores copied from the
MedleyVox paper (\protect\hyperlink{ref-medleyvox}{Jeon et al., 2023}).

\begin{longtable}[]{@{}lcc@{}}
\toprule\noalign{}
Methods & SI-SDRi & SDRi \\
\midrule\noalign{}
\endhead
\bottomrule\noalign{}
\endlastfoot
iSRNet (\protect\hyperlink{ref-medleyvox}{Jeon et al., 2023}) & 15.10 &
14.20 \\
NMF & 5.12 & 5.97 \\
Naive & 6.61 \(\pm\) 0.25 & 7.60 \(\pm\) 0.21 \\
Segmented & 11.14 \(\pm\) 0.48 & 11.77 \(\pm\) 0.47 \\
AR (proposed) & \textbf{11.24} \(\pm\) 0.40 & \textbf{11.89} \(\pm\)
0.34 \\
AR w/ TF & 11.75 \(\pm\) 0.38 & 12.34 \(\pm\) 0.39 \\
\end{longtable}

All the proposed methods outperformed the learning-free NMF baseline.
Both the auto-regressive sampling and the segmented sampling improved
the naive sampling baseline. The reason is because both methods operate
on a smaller segment, thus giving finer control to pick the best sample.

Comparing the auto-regressive and segmented sampling, the effect of the
extra information does give AR sampling slightly higher scores and less
variance. However, the difference is not significant considering we only
performed three runs for calculating the statistics. To examine whether
the condition plays a role in the sampling process, we also performed
sampling with teacher forcing, i.e.~using the ground truth signal of the
overlapping part as condition. We name this variant as AR w/ TF. The
results in the table show that accurate condition improves around 0.6 dB
in both metrics.

\hypertarget{conclusions}{%
\section{Conclusions}\label{conclusions}}

In this paper, we examined the problem of using an unconditional
diffusion model to separate monotimbral sources, in the context of duet
singing voices separation. We proposed sequentially sampling on the
overlapping mixture segments to enforce the coherency of singer identity
in the separated audio. We evaluated the proposed method on the
MedleyVox dataset. The results show that, although most of the
improvements are because of the finer control of the sampling process,
conditioning on the previous segment does improve the separation
performance over the naive posterior sampling baseline.

\hypertarget{references}{%
\section*{References}\label{references}}
\addcontentsline{toc}{section}{References}

\hypertarget{refs}{}
\begin{CSLReferences}{1}{0}
\leavevmode\vadjust pre{\hypertarget{ref-csd}{}}%
Choi, S., Kim, W., Park, S., Yong, S., \& Nam, J. (2020). Children's
song dataset for singing voice research. \emph{International Society for
Music Information Retrieval Conference (ISMIR)}.

\leavevmode\vadjust pre{\hypertarget{ref-jukebox}{}}%
Dhariwal, P., Jun, H., Payne, C., Kim, J. W., Radford, A., \& Sutskever,
I. (2020). Jukebox: A generative model for music. \emph{arXiv Preprint
arXiv:2005.00341}.

\leavevmode\vadjust pre{\hypertarget{ref-nus}{}}%
Duan, Z., Fang, H., Li, B., Sim, K. C., \& Wang, Y. (2013). The NUS sung
and spoken lyrics corpus: A quantitative comparison of singing and
speech. \emph{2013 Asia-Pacific Signal and Information Processing
Association Annual Summit and Conference}, 1--9.

\leavevmode\vadjust pre{\hypertarget{ref-vrdmg}{}}%
Hernandez-Olivan, C., Saito, K., Murata, N., Lai, C.-H.,
Mart{\'\i}nez-Ramirez, M. A., Liao, W.-H., \& Mitsufuji, Y. (2023). {VRDMG}:
Vocal restoration via diffusion posterior sampling with multiple
guidance. \emph{arXiv Preprint arXiv:2309.06934}.

\leavevmode\vadjust pre{\hypertarget{ref-hirano2023diffusion}{}}%
Hirano, M., Kazuki, S., Koyama, Y., Takahashi, S., \& Mitsufuji, Y.
(2023). Diffusion-based signal refiner for speech enhancement.
\emph{arXiv Preprint arXiv:2305.05857}.

\leavevmode\vadjust pre{\hypertarget{ref-ddpm}{}}%
Ho, J., Jain, A., \& Abbeel, P. (2020). Denoising diffusion
probabilistic models. \emph{Advances in Neural Information Processing
Systems}, \emph{33}, 6840--6851.

\leavevmode\vadjust pre{\hypertarget{ref-opensinger}{}}%
Huang, R., Chen, F., Ren, Y., Liu, J., Cui, C., \& Zhao, Z. (2021).
Multi-singer: Fast multi-singer singing voice vocoder with a large-scale
corpus. \emph{Proceedings of the 29th ACM International Conference on
Multimedia}, 3945--3954.

\leavevmode\vadjust pre{\hypertarget{ref-medleyvox}{}}%
Jeon, C.-B., Moon, H., Choi, K., Chon, B. S., \& Lee, K. (2023).
{MedleyVox}: An evaluation dataset for multiple singing voices
separation. \emph{ICASSP 2023-2023 IEEE International Conference on
Acoustics, Speech and Signal Processing (ICASSP)}, 1--5.

\leavevmode\vadjust pre{\hypertarget{ref-ddrm}{}}%
Kawar, B., Elad, M., Ermon, S., \& Song, J. (2022). Denoising diffusion
restoration models. \emph{Advances in Neural Information Processing
Systems}, \emph{35}, 23593--23606.

\leavevmode\vadjust pre{\hypertarget{ref-pjs}{}}%
Koguchi, J., Takamichi, S., \& Morise, M. (2020). {PJS}:
Phoneme-balanced japanese singing-voice corpus. \emph{2020 Asia-Pacific
Signal and Information Processing Association Annual Summit and
Conference (APSIPA ASC)}, 487--491.

\leavevmode\vadjust pre{\hypertarget{ref-adamw}{}}%
Loshchilov, I., \& Hutter, F. (2017). Decoupled weight decay
regularization. \emph{arXiv Preprint arXiv:1711.05101}.

\leavevmode\vadjust pre{\hypertarget{ref-manilow2021unsupervised}{}}%
Manilow, E., O'Reilly, P., Seetharaman, P., \& Pardo, B. (2021).
Unsupervised source separation by steering pretrained music models.
\emph{arXiv Preprint arXiv:2110.13071}.

\leavevmode\vadjust pre{\hypertarget{ref-mariani2023multi}{}}%
Mariani, G., Tallini, I., Postolache, E., Mancusi, M., Cosmo, L., \&
Rodolà, E. (2023). Multi-source diffusion models for simultaneous music
generation and separation. \emph{arXiv Preprint arXiv:2302.02257}.

\leavevmode\vadjust pre{\hypertarget{ref-moliner2023solving}{}}%
Moliner, E., Lehtinen, J., \& Välimäki, V. (2023). Solving audio inverse
problems with a diffusion model. \emph{ICASSP 2023-2023 IEEE
International Conference on Acoustics, Speech and Signal Processing
(ICASSP)}, 1--5.

\leavevmode\vadjust pre{\hypertarget{ref-murata2023gibbsddrm}{}}%
Murata, N., Saito, K., Lai, C.-H., Takida, Y., Uesaka, T., Mitsufuji,
Y., \& Ermon, S. (2023). {GibbsDDRM}: A partially collapsed gibbs
sampler for solving blind inverse problems with denoising diffusion
restoration. \emph{arXiv Preprint arXiv:2301.12686}.

\leavevmode\vadjust pre{\hypertarget{ref-asteroid}{}}%
Pariente, M., Cornell, S., Cosentino, J., Sivasankaran, S., Tzinis, E.,
Heitkaemper, J., Olvera, M., Stöter, F.-R., Hu, M., Martín-Doñas, J. M.,
Ditter, D., Frank, A., Deleforge, A., \& Vincent, E. (2020). Asteroid:
The {PyTorch}-based audio source separation toolkit for researchers.
\emph{Proc. Interspeech}.

\leavevmode\vadjust pre{\hypertarget{ref-crash}{}}%
Rouard, S., \& Hadjeres, G. (2021). {CRASH}: Raw audio score-based
generative modeling for controllable high-resolution drum sound
synthesis. \emph{arXiv Preprint arXiv:2106.07431}.

\leavevmode\vadjust pre{\hypertarget{ref-saito2023unsupervised}{}}%
Saito, K., Murata, N., Uesaka, T., Lai, C.-H., Takida, Y., Fukui, T., \&
Mitsufuji, Y. (2023). Unsupervised vocal dereverberation with
diffusion-based generative models. \emph{ICASSP 2023-2023 IEEE
International Conference on Acoustics, Speech and Signal Processing
(ICASSP)}, 1--5.

\leavevmode\vadjust pre{\hypertarget{ref-mousai}{}}%
Schneider, F., Jin, Z., \& Schölkopf, B. (2023). Mo\(\backslash\)\^{}
usai: Text-to-music generation with long-context latent diffusion.
\emph{arXiv Preprint arXiv:2301.11757}.

\leavevmode\vadjust pre{\hypertarget{ref-schulze2023unsupervised}{}}%
Schulze-Forster, K., Richard, G., Kelley, L., Doire, C. S., \& Badeau,
R. (2023). Unsupervised music source separation using differentiable
parametric source models. \emph{IEEE/ACM Transactions on Audio, Speech,
and Language Processing}, \emph{31}, 1276--1289.

\leavevmode\vadjust pre{\hypertarget{ref-song2020score}{}}%
Song, Y., Sohl-Dickstein, J., Kingma, D. P., Kumar, A., Ermon, S., \&
Poole, B. (2020). Score-based generative modeling through stochastic
differential equations. \emph{arXiv Preprint arXiv:2011.13456}.

\leavevmode\vadjust pre{\hypertarget{ref-jvs}{}}%
Tamaru, H., Takamichi, S., Tanji, N., \& Saruwatari, H. (2020).
{JVS-MuSiC}: Japanese multispeaker singing-voice corpus. \emph{arXiv
Preprint arXiv:2001.07044}.

\leavevmode\vadjust pre{\hypertarget{ref-opencpop}{}}%
Wang, Y., Wang, X., Zhu, P., Wu, J., Li, H., Xue, H., Zhang, Y., Xie,
L., \& Bi, M. (2022). Opencpop: A high-quality open source chinese
popular song corpus for singing voice synthesis. \emph{arXiv Preprint
arXiv:2201.07429}.

\leavevmode\vadjust pre{\hypertarget{ref-vocalset}{}}%
Wilkins, J., Seetharaman, P., Wahl, A., \& Pardo, B. (2018). {VocalSet}:
A singing voice dataset. \emph{ISMIR}, 468--474.

\leavevmode\vadjust pre{\hypertarget{ref-yu2018multi}{}}%
Yu, C.-Y., \& Su, L. (2018). Multi-layered cepstrum for instantaneous
frequency estimation. \emph{2018 IEEE Global Conference on Signal and
Information Processing (GlobalSIP)}, 276--280.

\leavevmode\vadjust pre{\hypertarget{ref-yu2023conditioning}{}}%
Yu, C.-Y., Yeh, S.-L., Fazekas, G., \& Tang, H. (2023). Conditioning and
sampling in variational diffusion models for speech super-resolution.
\emph{ICASSP 2023-2023 IEEE International Conference on Acoustics,
Speech and Signal Processing (ICASSP)}, 1--5.

\leavevmode\vadjust pre{\hypertarget{ref-zai2022transfer}{}}%
Zai El Amri, W., Tautz, O., Ritter, H., \& Melnik, A. (2022). Transfer
learning with jukebox for music source separation. \emph{IFIP
International Conference on Artificial Intelligence Applications and
Innovations}, 426--433.

\leavevmode\vadjust pre{\hypertarget{ref-m4singer}{}}%
Zhang, L., Li, R., Wang, S., Deng, L., Liu, J., Ren, Y., He, J., Huang,
R., Zhu, J., Chen, X., \& others. (2022). {M4Singer}: A multi-style,
multi-singer and musical score provided mandarin singing corpus.
\emph{Advances in Neural Information Processing Systems}, \emph{35},
6914--6926.

\end{CSLReferences}

\end{document}